\documentclass[RNAAS]{aastex62}

\begin{document}

\title{ASAS-SN Identification of a detached eclipsing binary system with a $\sim7.3$ year period}

\correspondingauthor{T. Jayasinghe}
\email{jayasinghearachchilage.1@osu.edu}

\author[0000-0002-6244-477X]{T. Jayasinghe}
\affiliation{Department of Astronomy, The Ohio State University, 140 West 18th Avenue, Columbus, OH 43210, USA}
\author{K. Z.  Stanek}
\affiliation{Department of Astronomy, The Ohio State University, 140 West 18th Avenue, Columbus, OH 43210, USA}
\affiliation{Center for Cosmology and Astroparticle Physics, The Ohio
State University, 191 W. Woodruff Avenue, Columbus, OH
43210}
\author{C. S. Kochanek}
\affiliation{Department of Astronomy, The Ohio State University, 140 West 18th Avenue, Columbus, OH 43210, USA}
\affiliation{Center for Cosmology and Astroparticle Physics, The Ohio
State University, 191 W. Woodruff Avenue, Columbus, OH
43210}
\author{B. J. Shappee}
\affiliation{Institute for Astronomy, University of Hawai’i, 2680 Woodlawn Drive, Honolulu, HI 96822,USA}

\author{T. W. -S. Holoien}
\affiliation{Carnegie Fellow, The Observatories of the Carnegie Institution for Science, 813 Santa Barbara St., Pasadena, CA 91101, USA}

\author{T. A. Thompson}
\affiliation{Department of Astronomy, The Ohio State University, 140 West 18th Avenue, Columbus, OH 43210, USA}
\affiliation{Center for Cosmology and Astroparticle Physics, The Ohio
State University, 191 W. Woodruff Avenue, Columbus, OH
43210}
\author{J. L. Prieto}
\affiliation{N\'ucleo de Astronom\'ia de la Facultad de Ingenier\'ia, Universidad Diego Portales, Av. Ej\'ercito 441, Santiago, Chile}
\affiliation{Millennium Institute of Astrophysics, Santiago, Chile}
\author{Subo Dong}
\affiliation{Kavli Institute for Astronomy and Astrophysics, Peking
University, Yi He Yuan Road 5, Hai Dian District, China}
\author{D. J.  Stevens}
\affiliation{Department of Astronomy, The Ohio State University, 140 West 18th Avenue, Columbus, OH 43210, USA}


\keywords{binaries: eclipsing --- stars: variables: general --- surveys}

\section{}
The All-Sky Automated Survey for SuperNovae (ASAS-SN, \citealt{2014ApJ...788...48S, 2017PASP..129j4502K}) routinely monitors the entire visible sky to a depth of $\sim17$ mag in both the V-band and g-band. Our cadence and all-sky coverage facilitates the study of a variety of transient and variable phenomena, including the discovery of supernovae, cataclysmic variables and variable stars. While the primary focus is on transients (e.g., \citealt{2017MNRAS.471.4966H}), we also published our first catalog of variable stars with over $\sim66,000$ new variables \citep{2018MNRAS.477.3145J}. As part of an effort to release V-band light curves for all known variables within the limits of our survey, we are uniformly analyzing the light curves of over $\sim 200,000$ known variables listed in the VSX catalog \citep{2006SASS...25...47W}. Using our automated variable star classification pipeline, we have identified numerous variables that require different classifications and periods, with the detailed results to appear in a future publication. 

\begin{figure}[htpb!]
\begin{center}
\includegraphics[scale=0.7,angle=0]{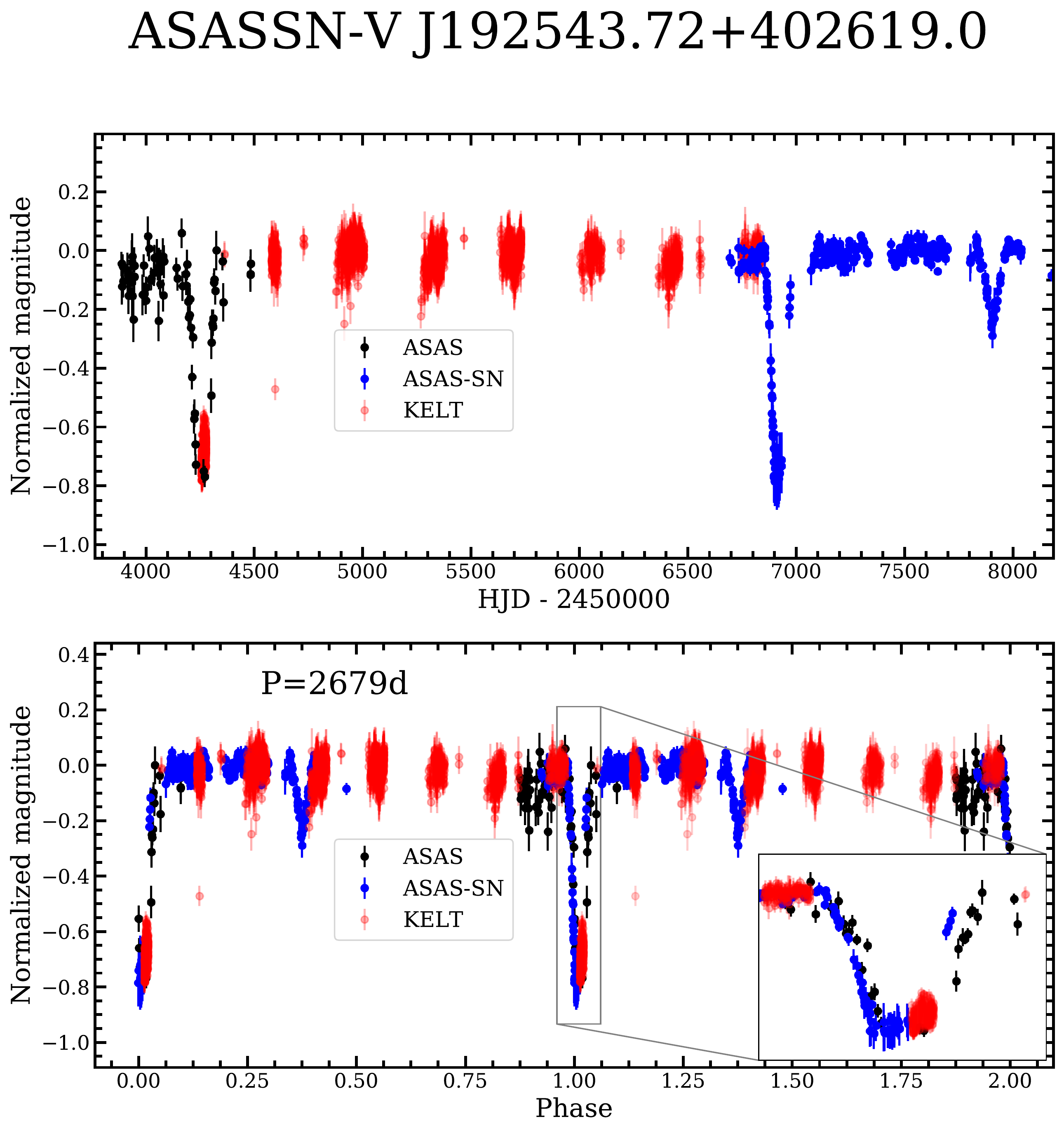}
\caption{The top panel shows the combined light curve for ASASSN-V J192543.72+402619.0 and the bottom panel shows the combined light curve phased with the optimum period of $P\sim2679$ d. Different colored points correspond to data from the different surveys.\label{fig:1}}
\end{center}
\end{figure}

Here we report the identification of a new, long period detached eclipsing binary system. ASASSN-V J192543.72+402619.0 (KIC 5273762) was discovered by the All-Sky Automated Survey (ASAS; \citealt{2002AcA....52..397P,2009AcA....59...33P}) as an aperiodic/irregular variable (ASAS J192544+4026.4), and is also part of the Kepler Input Catalog \citep{2011AJ....142..112B}. The ASAS light curve captures one primary eclipse, but lacked sufficient phase coverage to identify it as an eclipsing binary. The ASAS-SN data spanning 2014-2018 showed evidence of both a primary and secondary eclipse, with the very long period triggering our interest. To confirm the classification and derive an accurate period, we supplement the ASAS-SN light curve with V-band data from the ASAS \citep{2002AcA....52..397P} and KELT surveys \citep{2007PASP..119..923P}, extending the light curve back to 2006 (Figure \ref{fig:1}). Following the prescription detailed in \citet{2018MNRAS.477.3145J}, we calculated periodograms and arrived at an orbital period of $P\sim 2679$ d. Eclipsing binaries with such periods are exceedingly rare, placing ASASSN-V J192543.72+402619.0 into a class of unusual binaries that includes TYC 2505-672-1 \citep{2016AJ....151..123R}, $\epsilon$ Aurigae \citep{2010Natur.464..870K} and EE Cephei \citep{2003A&A...403.1089G} that have periods of 69.1 yr, 27.1 yr and 5.6 yr respectively. The light curve for ASASSN-V J192543.72+402619.0 also shows a secondary eclipse that differs in depth from the primary eclipse by $\sim0.6$ mag, which is unusual for this class of long period binaries.

Multi-band photometry from WISE \citep{2010AJ....140.1868W} and 2MASS \citep{2006AJ....131.1163S} gives the colors $W1-W2=-0.17$ and $J-K_s=1.16$, suggesting that at least one component of this binary is an M-giant \citep{2014EAS....67..409Z}. We obtain a distance of $D\sim 5$ kpc using the Gaia DR2 parallax \citep{2018arXiv180409365G} giving $M_{K_S}=-6.4$, which is typical of a M-giant. The primary eclipse profile is asymmetric and is similar to the profiles seen in disk-eclipse binaries \citep{2014ApJ...788...41D}. The primary eclipse lasts $\sim160$ days, and the next eclipse is expected to begin in November-December, 2021. Using the procedure detailed by \citet{2013ApJ...763L...2D}, we calculated the minimum orbital eccentricity to be $e_{min}\sim0.2$. Based on these lines of evidence, it is likely that ASASSN-V J192543.72+402619.0 is another disk eclipsing binary. We encourage further photometric and spectroscopic observations to discern the true nature of this system. A full analysis of the existing data is in preparation (Dixon et al. 2018).

\bibliographystyle{plainnat}

\end{document}